**Bashmur K.A., Kachaeva V.A., Bukhtoyarov V.V., Saramud M.V.**
# MAGNETOREOLOGICAL SPRING AS ELEMENT OF VIBRATION CONTROL SYSTEM OF DYNAMICLY ACTIVE EQUIPMENT FOR BIOFUEL PRODUCTION
Siberian Federal University, Krasnoyarsk

Abstract: The development of vibration protection systems that ensure efficiency and safety in the operation of process equipment and pipelines is one of the main tasks of controlling the dynamic state of machines. One of the effective methods of vibration isolation of the equipment of these installations is the use of vibration mounts. Today, both passive and active methods of extinguishing are actively used. The combination of two methods of damping will ensure the stability and adaptability of vibration protection systems to the operating conditions of process equipment. The paper presents and investigates the device of a hybrid vibration support, including a magnetorheological spring, as an element of vibration damping. A distinctive feature of the vibration mount is an extended range of performance and reduced wear rate of the components. An analysis of the damping characteristics of a hybrid vibration support in passive and active modes of operation is presented. A simulation model of the vibration support under consideration has been developed in the MATLAB Simulink simulation environment using the PIB controller. The simulation results indicate the effectiveness of the use of vibration damping systems with a magnetorheological spring in conjunction with various technological *equipment*.

*Башмур К.А., Качаева В.А., Бухтояров В.В., Сарамуд М.В.*
**МАГНИТОРЕОЛОГИЧЕСКАЯ ПРУЖИНА КАК ЭЛЕМЕНТ СИСТЕМЫ УПРАВЛЕНИЯ ВИБРАЦИЯМИ ДИНАМИЧЕСКИ АКТИВНОГО ОБОРУДОВАНИЯ ДЛЯ ПРОИЗВОДСТВА БИОТОПЛИВА**
Сибирский федеральный университет, Красноярск

**Аннотация:** Разработка виброзащитных систем, обеспечивающих эффективность и безопасность при работе технологического оборудования и трубопроводов, одна из основных задач управления динамическим состоянием машин. Одним из эффективных методов виброизоляции оборудования данных установок является применение виброопор. На сегодняшний день активно применяются как пассивные, так и активные способы гашения. Совмещение двух способов гашения позволит обеспечить стабильность и адаптивность виброзащитных систем к условиям работы технологического оборудования. В работе представлено и исследовано устройство гибридной виброопоры, включающей магнитореалогическую пружину, как элемент виброгашения. Отличительной чертой виброопоры является расширенный диапазон рабочих характеристик и уменьшенная интенсивность износа составных частей. Представлен анализ демпфирующих характеристик гибридной виброопоры при пассивном и активном режиме работы. Разработана имитационная модель рассматриваемой виброопоры в среде имитационного моделирования MATLAB Simulink с использованием PIB регулятора. Результаты моделирования свидетельствуют об эффективности применения систем виброгашения с магнитореологической пружиной совместно с различным технологическим оборудованием.

**Введение**

Вибрационная защита оборудования обеспечивает безопасность и надежность работы техники, обслуживающего персонала и окружающей среды в целом. Разработка способов защиты от вибрации основная задача управления динамическим состоянием машин. В результате современные виброзащитные системы становятся специализированными системами автоматического управления.

Распространено гашение колебаний оборудования за счет использование пассивных систем виброгашения. Пассивное виброгашение происходит только за счет использования упругодемпфирующих элементов, например, пружин. Они имеют предписанную жесткость и обычно небольшой коэффициент демпфирования [1]. Упругодемпфирующие элементы интенсивно изнашиваются, что повышает риск выхода оборудования из строя. Пассивное виброгашение происходит только в ограниченном диапазоне характеристик.

Активное виброгашение предполагает возможность непрерывного управления демпфирующими характеристиками, например, с помощью магнитореологической жидкости [2]. На данный момент применение исполнительных элементов на основе магнитореологической жидкости заложило основы нового этапа в развитии виброзащитных систем. Это позволяет виброзащитным устройствам адаптироваться к параметрам работы технологического оборудования.

Активное виброгашение имеет ряд преимуществ перед пассивным режимом: широкий диапазон рабочих характеристик, адаптивность, надежность, уменьшение износа составных частей [3,4]. Коэффициент демпфирования может быть изменен в системах с магнитореологической пружиной, что позволит использовать преимущества активного виброгашения. Но такие типы конструкций обладают рядом недостатков, такими как: низкая надежность в связи с возможным запаздыванием показаний датчиков и управляющих устройств; функциональные ограничения в возможности применения ввиду зависимости от источников электроэнергии [5]. В результате совместное использование упругодемпифрующих элементов и системы магнитореологической пружины позволит безопасно эксплуатировать динамически активное оборудование и трубопроводы.

**Теория**

Анализируя преимущества и недостатки способов виброгашения, на базе Лаборатории биотопливных композиций Сибирского федерального университета предложена гибридная конструкция виброопоры на основе виброопоры со сферическими упругодемпфирующими элементами [6]. Гибридная виброопора совмещает в себе пассивное и активное виброгашение. Это позволяет повысить эффективность виброгашения за счет расширения рабочих характеристик и повышения надежности. Конструкция гибридной виброопоры представлена на рисунке 1.

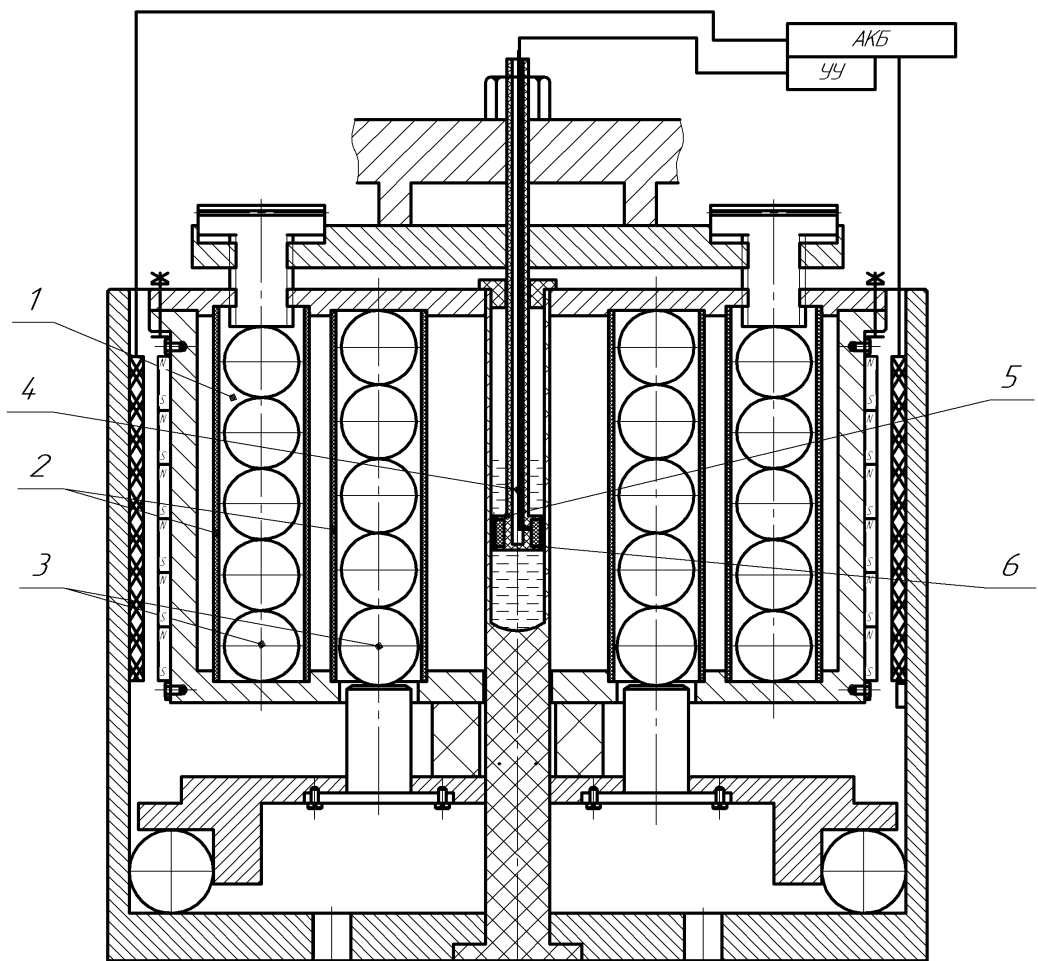

Рис. 1 – Конструкция полуактивной виброопоры : 1 – упругоразгрузочные модули; 2 – направляющие трубки; 3 – упругодемпфирующие элементы; 4 – шток; 5 – поршень; 6 – электромагнит

Гибридная виброопора содержит упругие модели 1, состоящие из направляющих трубок 2 и упруго-разгрузочных элементов 3, размещенных в один ряд. Эластичные модели действуют как пассивная виброизоляция. Полость штока содержит магнитореологическую систему, позволяющую плавно регулировать коэффициент демпфирования гибридной виброопоры. Активная система состоит из полости, заполненной магнитореологической жидкостью. Полость содержит шток 4 и поршень 5 с электромагнитом 6. Поршень имеет дроссельные каналы, обеспечивающие перетекание магнитореологической жидкости из одной гидрополости в другую.

Система работает следующим образом. Приложенное к штоку усилие на режиме «сжатие» заставляет перемещаться поршень с обмоткой вдоль центральной оси по направлению к нижнему основанию. Это сопровождается перетеканием магнитореологической жидкости через дроссельные каналы поршня МР-демпфера из одной полости в другую. Кавитация магнитореологической жидкости, возникающая при резких перемещениях поршня МР-демпфера, устраняется поджатием магнитореологической жидкости через поршень газом, закачанным под давлением. Подвод напряжения к обмотке электромагнита осуществляется через контакты, выведенные наружу через отверстие в крышке. Степень вязкости магнитореологической жидкости регулируется с помощью тока, протекающего по обмотке электромагнита, чем достигается изменение силы гидравлического сопротивления.

Управление силой гидравлического сопротивления магнитореологической системы осуществляется посредством устройства управления (УУ), выполненного на аналоговых или микропроцессорных элементах. При управлении гидравлическим сопротивлением УУ может учитывать информацию, поступающую от датчиков ускорения защищаемого объекта и опорной поверхности. Сигнал управления с выхода УУ поступает на выходной зажим катушки электромагнита МР-система. Аккумулятор служит для накопления вырабатываемой электроэнергии, необходимой для питания МР-система и другого сопутствующего оборудования.

**Результаты эксперимента**
**Расчетная схема**

Несмотря на то, что сложные технические объекты имеют сложные расчетные схемы в виде колебательных систем с несколькими степенями свободы, современные научные исследования связаны с упрощением исходных систем с учетом специфики динамических взаимодействий элементов [7]. В качестве расчетных схем в динамике машин наибольшее распространение получили механические колебательные системы с одной, двумя или тремя степенями свободы. Именно такие системы являются базовыми. Основные упрощения связаны с допущением, что объект защиты и агрегаты представляются собой материальными точками [8].

В случае виброизолирующего оборудования зачастую используется расчетная схема двухмассовой колебательной системы. Расчетная схема двухмассовой колебательной системы с двумя степенями свободы представлена на рисунке 2, где символом $m_1$ обозначена неподрессоренная масса, являющаяся промежуточным объектом виброзащиты и защищенная от возмущения $Z_0$ упругим элементом с жесткостью $C_1$ и элементом гидравлического сопротивления с коэффициентом $\beta_1$. Символом $m_2$ обозначена подрессоренная масса, дополнительно защищенная от возмущающего воздействия $Z_0$ со стороны основания упругим элементом с жесткостью $C_2$ и элементом гидравлического сопротивления с коэффициентом $\beta_2$.

Подрессоренная $m_2$ и неподрессоренная $m_1$ массы имеют координаты перемещения $Z_1$ и $Z_2$ соответственно. Коэффициент гидравлического сопротивления $\beta_2$ зависит от конструкции демпфера, определяющей его постоянное значение, а также тока в его катушке, определяющего его изменяемое значение.

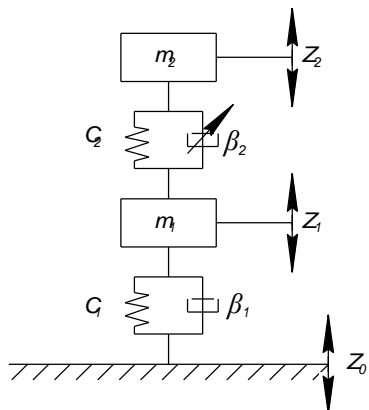

Рисунок 2 - Расчетная схема двумассовой системы виброзащиты

В активной системе виброзащиты управление током в обмотке МР-демпфера осуществляется в функции координат виброзащищаемого объекта, то есть значение коэффициента гидравлического сопротивления изменяется в процессе управления. Для расчетной схемы устанавливается зависимость между регулирующим воздействием в виде тока в катушке магнитореологического демпфера и регулируемой переменной в виде перемещения, виброскорости и виброускорения защищаемого объекта. Управление такое системой называют параметрическим, то есть управление жесткостью или коэффициентом гидравлического сопротивления элементов системы виброзащиты [10].

Уравнение движения конструкции записывается в виде:
$$\begin{cases} m_2 \cdot \ddot{z}_2 + f_2 = 0, \\ m_1 \cdot \ddot{z}_1 + f_1 - f_2 = 0, \end{cases}$$
где $f_2$, $f_1$ – сила между объектом, имеющим массу $m_2$, и демпфером 2, сила между $m_1$ и демпфером 1, $m_2$, $m_1$ – масса объектов, $z_1$, $z_2$ – вектор смещения конструкции, перемещения объекта, чем имеют массу $m_1$, перемещение объекта, чем у массы $m_2$.

Сила между объектом массой $m_2$ и демпфером 2 моделировалась с помощью пружины и демпфера. Сила может быть выражена как:
$$\begin{cases} f_2 = \beta_2 \cdot (\dot{z}_2 - \dot{z}_1) + C_2 \cdot (z_2 - z_1), z_2 - z_1 \geq 0 \\ f_2 = 0, z_2 - z_1 < 0 \end{cases},$$
где $\beta_2$ и $C_2$ — коэффициенты демпфирования и жесткости.

Аналогично с помощью пружины и демпфера моделировалась сила между объектом, имеющим массу $m_1$, и демпфером 1. Сила может быть выражена как:
$$\begin{cases} f_1 = \beta_1 \cdot (\dot{z}_1 - \dot{z}_0) + C_1 \cdot (z_1 - z_0), z_1 - z_0 \geq 0 \\ f_1 = 0, z_1 - z_0 < 0 \end{cases},$$
где $\beta_1$ и $C_1$ – коэффициенты демпфирования и жесткости, $z_1$, $z_0$ – водоизмещающий объект, имеющий массу $m_1$, и основание.

В результате получаем следующую математическую модель:
$$\begin{cases} m_2 \cdot \ddot{z}_2 = -\beta_2 \cdot (\dot{z}_2 - \dot{z}_1) - C_2 \cdot (z_2 - z_1) \\ m_1 \cdot \ddot{z}_1 = -\beta_2 \cdot (\dot{z}_2 - \dot{z}_1) - C_2 \cdot (z_2 - z_1) - \\ \quad - \beta_1 \cdot (\dot{z}_1 - \dot{z}_0) - C_1 \cdot (z_1 - z_0) \end{cases} \quad (1)$$

**Создание имитационной модели**

На основе математической модели (1) проанализируем демпфирующие характеристики системы. На основании системы уравнений (1) построена имитационная модель (рис. 3) в графической среде имитационного моделирования MATLAB Simulink, что обусловлено наличием проверенных математических методов, наглядностью и точностью представления результатов моделирования. Возмущающее воздействие может задаваться различными законами, в том числе и случайно, когда воздействие определяется MATLAB автоматически в заданных пределах. Использование такого воздействия в системе позволяет добиться максимального приближения теоретического исследования к производственному процессу.

Правая часть уравнения (1) составляется следующим образом. Коэффициент демпфирования умножается на разницу первых производных перемещения виброизолируемого объекта и возмущающего воздействия. Общий коэффициент упругости умножается на разницу перемещения виброизолируемого объекта и возмущающего воздействия. Для выполнения данных операция применяются блоки Product. Далее оба произведения умножаются на коэффициент -1 и складываются между собой с помощью блока Add. После деления полученной суммы на массу виброзащитной системы создаем связь с ускорением виброизолируемого объекта. Составление уравнения (11) происходит аналогично. В результате полученная схема является полностью замкнутой. Коэффициент упругости системы получаем с использованием законов физики, исходя из последовательного или параллельного соединения элементов. Для большей наглядности и получения количественной характеристики установлен дисплей, отражающей сумму перемещение технологической установки относительно фундамента. На дисплей выводится разница между полученным сигналом и желаемым. В данном случае желаемое перемещение системы равно 0.

Система на рисунке 3 построена с учетом численных значений параметров, соответствующих разработанной экспериментальной установке для исследования динамических характеристик системы виброзащиты с магнитореологическим демпфером колебаний, при этом коэффициент демпфирования k остается постоянным, что соответствует пассивному способу виброгашения. В результате запуска системы получен график колебаний, представленный на рисунке 4.

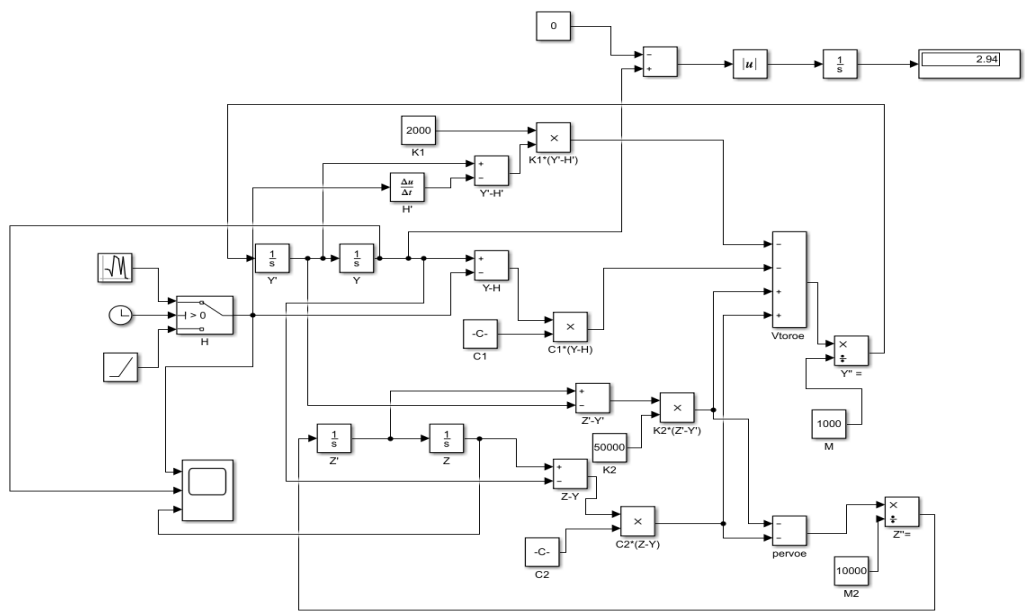

Рисунок 3 - Программа для анализа колебаний системы при отсутствии системы регулирования

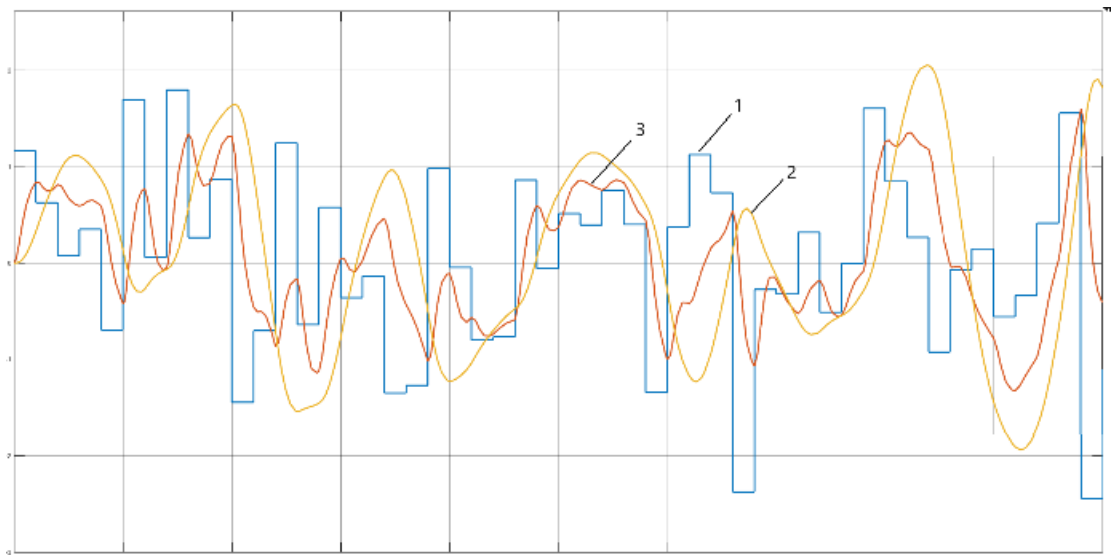

Рисунок 4 – График колебаний частей системы от времени: 1 – возмущающее воздействие, 2 – гибридная виброопора, 3- виброизолируемый объект

Анализ системы показывает, что даже при небольшом возмущающем воздействии, система приходит в движение, несмотря на наличие пассивной виброизоляции. Стоит отметить, что при выбранном возмущающем воздействии колебания находятся в пределе допустимых значений, значит активная виброопора справляется с основной задачей. Но в случае увеличения возмущающего воздействия, увеличится и ответное перемещение системы, что, соответственно, повышает риск выхода из строя оборудования, получения травм персоналу и нанесение ущерба окружающей среде при самом неблагоприятном исходе.

В результате запуска системы общее перемещение компрессорной установки составляет 2.94 у.е. Данное перемещение свидетельствует о больших вибрациях, возникающих при работе оборудования. Это свидетельствует о неэффективном виброгашении системой. Изменение характеристик колебательного процесса свидетельствует о том, что при пассивной работе виброопора не может подстраиваться под возмущающее воздействие.

Далее рассмотрим систему виброзащиты при активном режиме работы, то есть в присутствии устройства управления (рис. 5), позволяющего регулировать коэффициент демпфирования. Необходимо изменить исходную схему. В данном случае коэффициент демпфирования, равный константе, заменяется на PID – регулятор. Использование PID-регуляторов в промышленности получило широкое распространение. Перед регулятором происходит сравнение желаемого значения перемещения виброизолируемого объекта и полученного виброперемещения. Настройка параметров PID-регулятора осуществляется в среде имитационного моделирования MATLAB Simulink. MATLAB автоматически подбирает параметры, при которых коэффициенты усиления пропорциональной, интегральной и дифференциальной составляющих оптимальны. В результате моделирования были получены графики виброперемещения виброизолированного объекта (рис. 6)

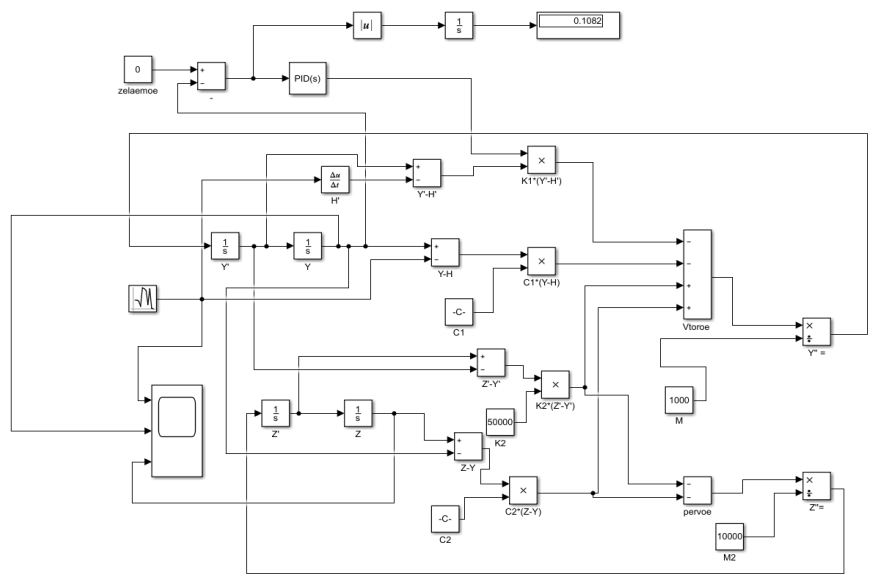

Рисунок 5 - Программа для анализа колебаний системы с устройством управления

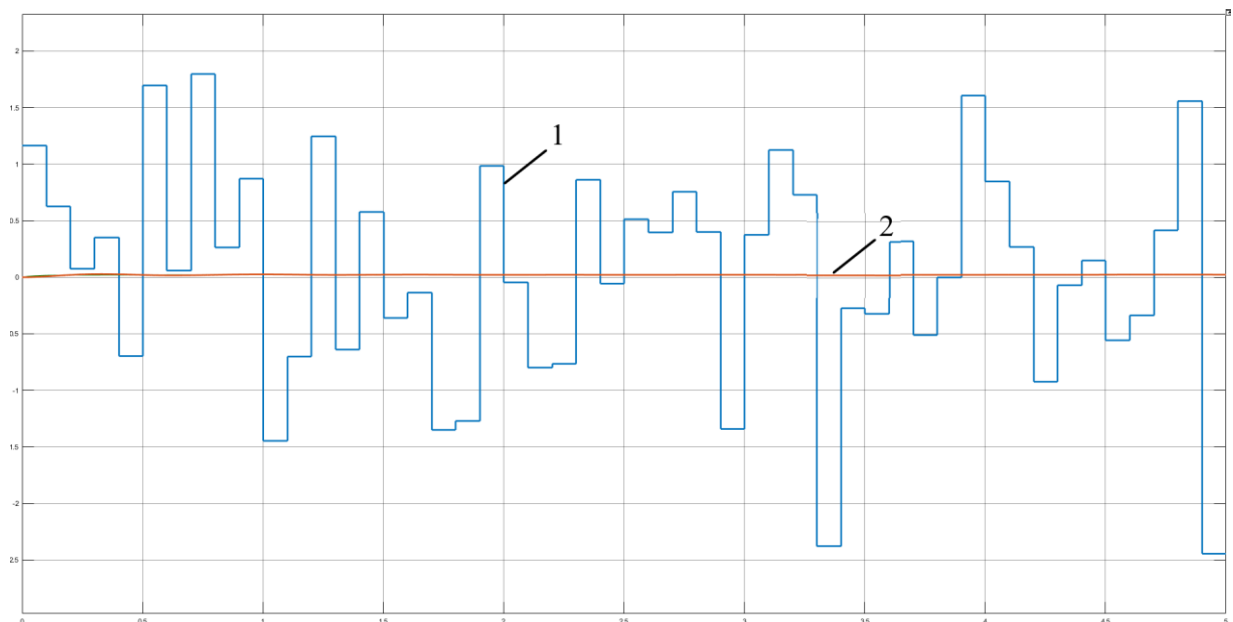

Рисунок 6 – График колебаний частей системы от времени: 1 – возмущающее воздействие, 2 – виброизолируемый объект

В результате общее перемещение виброизолируемого объекта за интервал времени равный 5 секундам и случайном возмущающем воздействии составляет 0.1062, что меньше общего перемещения при отсутствии регулятора. Перемещение виброизолируемого объекта в любой момент времени в течение срока испытаний не превышает 0.1. В результате можно сделать вывод, что система быстро реагирует на возмущающее воздействие и эффективно выполняет, поставленную задачу. Полученные результаты свидетельствуют об эффективности работы виброгасящей системы.

При сравнении графиков, представленных на рисунках 4 и 6, можно сделать вывод, что визуальное представление виброперемещений меньше при использовании системы с настроенным регулятором. Это доказывает эффективность применения активного режима работы виброопоры, то есть использование МР-демпфера. Несмотря на быстро изменяющееся возмущающее воздействие, системы подстраивается под него за счет изменение коэффициента демпфирования. В результате колебания технологического объекта минимальны, что позволит эксплуатировать оборудования без риска выхода из строя и аварий.

**Выводы**

В связи с тем, что вибрации являются сопутствующим фактором работы любого динамического объекта, существует необходимость совершенствования систем виброгашения. В результате анализа активного и пассивного способа гашения вибраций технологического оборудования разработано гибридная конструкция виброопоры. Такая виброопора совмещает в себе преимущества пассивного и активного виброгашения.

На основе математической модели построена имитационная модель виброгасящей системы в графической среде имитационного моделирования MATLAB Simulink при пассивном и активном режиме работы. Получены и приведены графики перемещений виброзолируемого объекта для пассивного и активного способа гашения колебаний. В результате имитационного моделирования суммарное перемещение виброизолируемого объекта в присутствии системы, изменяющей коэффциент дампирования меньше в 18 раз, чем при постоянном коэффициенте демпфирования.

В результате сравнения полученных результатов для двух вариантов работы оборудования, можно прийти к заключению, что колебания технологической установки в присутствии активной виброопоры будут заметно снижены. Это позволит использовать такую систему на производстве с целью снижения негативного влияния вибрации на оборудование.

Сравнение результатов, полученных для двух вариантов работы оборудования, показывает, что вибрация динамического оборудования при наличии магнитореологической системы будет заметно снижена. Это позволит использовать такую систему на производстве с целью снижения негативного воздействия вибрации на оборудование. Нельзя исключать и недостатки активного гашения вибрации, связанные со скоростью работы системы. Это ограничивает характеристики демпфирования и требует более детального анализа. Однако использование в производстве систем, сочетающих два способа гашения вибрации, позволяет получить более эффективные показатели.